\documentclass[showpacs,preprintnumbers,
amsmath,amssymb,aps,prd,nofootinbib,superscriptaddress,notitlepage,11pt]{revtex4-1}
\usepackage[T1]{fontenc}
\usepackage{amsmath,xcolor}
\usepackage{amsmath, amssymb,ascmac,fancybox,mathrsfs}
\usepackage{color,booktabs,fancyhdr,bm,comment,cancel}
\usepackage{hyperref,appendix}
\usepackage[]{graphicx,graphics}

\makeatletter
\renewcommand{\p@subsection}{}
\makeatother

\makeatletter
\newcommand{\xequal}[2][]{\ext@arrow 0055{\equalfill@}{#1}{#2}}
\def\equalfill@{\arrowfill@\Relbar\Relbar\Relbar}
\makeatother

\renewcommand{\thesection}{\arabic{section}}
\renewcommand{\thesubsection}{\arabic{section}.\arabic{subsection}}

\makeatletter
\renewcommand{\theequation}{\arabic{section}.\arabic{equation}}
\@addtoreset{equation}{section} 
\makeatother

\newcommand{\chushi}[1]{ }

\newcommand{\angleN}[1]{ \langle #1 \rangle }

\newcommand{\roundLR}[1]{ \left( #1 \right) }

\newcommand{\squareN}[1]{ [ #1 ] }

\newcommand{\absN}[1]{ | #1 | }

\let\calccommentout\iffalse 
\let\calcshow\iftrue 

\newcommand {\mathsym}[1]{{}}
\newcommand {\unicode}[1]{{}}

\begin{document}
\count\footins = 1000 

\title{A holographic study of the characteristics of chaos and correlation in the presence of backreaction}

\author{Shankhadeep Chakrabortty}
\email{s.chakrabortty@iitrpr.ac.in}
\affiliation{Department of Physics, Indian Institute of Technology Ropar, Rupnagar, Punjab 140 001, India}

\author{Hironori Hoshino}
\email{hhoshino.phys@gmail.com}
\affiliation{Department of Physics, Indian Institute of Technology Ropar, Rupnagar, Punjab 140 001, India}
\affiliation{Yukuhashi Tutoring School of Science and Mathematics, Nishimiyaichi 4-1-15, Yukuhashi, Fukuoka, 824-0033, Japan}

\author{Sanjay Pant}
\email{2018phz0012@iitrpr.ac.in}
\affiliation{Department of Physics, Indian Institute of Technology Ropar, Rupnagar, Punjab 140 001, India}

\author{Karunava Sil}
\email{ks45@iitbbs.ac.in}
\affiliation{School of Basic Sciences, Indian Institute of Technology Bhubaneswar, Bhubaneswar 752050, India}

\begin{abstract}

In this work, we perform a holographic study to estimate the effect of backreaction on the correlation between two subsystems forming the thermofield double (TFD) state. Each of these subsystems is described as a strongly coupled large-$N_c$ thermal field theory, and the backreaction imparted to it is sourced by the presence of a uniform distribution of heavy static quarks. The TFD state we consider here holographically corresponds to an entangled state of two AdS blackholes, each of which is deformed by a uniform distribution of static strings. In order to make a holographic estimation of correlation between two entangled boundary field theories in presence of backreaction we compute the holographic mutual information in the backreacted eternal blackhole. The late time exponential growth of an early perturbation is a signature of chaos in the boundary thermal field theory. Using the shock wave analysis in the dual bulk theory, we characterize this chaotic behaviour by computing the holographic butterfly velocity. We find that there is a reduction in the butterfly velocity due to a correction term that depends on the backreaction parameter. The late time exponential growth of an early perturbation destroys the two-sided correlation, whereas the backreaction always acts in favour of it. Finally we compute the entanglement velocity that essentially encodes the rate of disruption of correlation between two boundary theories. 
\end{abstract}

\maketitle
\section{Introduction}\label{S1}
\begin{figure}
\centering
  \includegraphics[width=.40\linewidth]{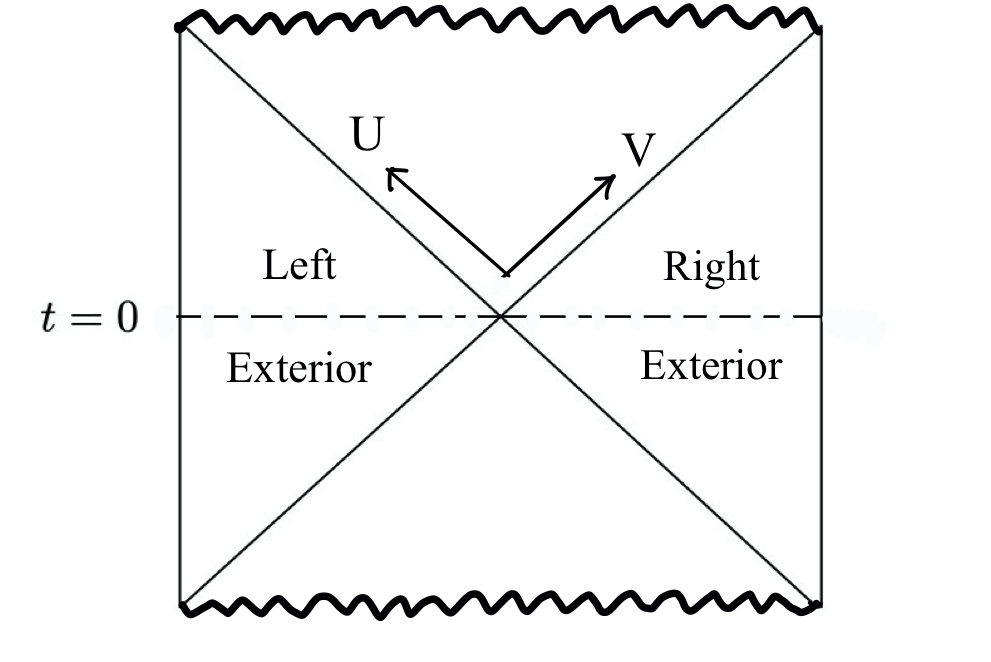}
  \caption{ Penrose diagram of the deformed eternal  blackhole. At t = 0, the spatial extremal surface connecting two asymptotic boundaries of eternal blackhole is denoted by the dashed line passing through the bifurcation point.
  }
\label{pen}
\end{figure}

In the context of AdS/CFT \cite {Maldacena:1997re, Witten:1998qj}, the thermofield double (TFD) state plays a crucial role in exploring the intriguing connection between black hole physics and quantum information theory.
 The holographic dual of TFD state turns out to be essential to diagnose the chaos and to study the information scrambling in the strongly coupled field theory \cite{Shenker:2013pqa, Shenker:2014cwa, Roberts:2014isa, Shenker:2013yza, Maldacena:2015waa,  Jahnke:2018off}. 
An example of TFD state can be presented as an entangled state, defined in a bipartite Hilbert space made out of the individual Hilbert spaces of two identically, non-interacting copies of large-$N_c$ strongly coupled thermal field theories.
Such a TFD state is holographically dual to two entangled AdS blackholes and the corresponding geometry is known as the eternal (two-sided) blackhole spacetime \cite{Israel:1976ur, Maldacena:2001kr}. In the Penrose diagram (figure \ref {pen}), the exterior region on both sides of the two-sided blackhole consists of two causally disconnected regions of spacetime, viz., the right ($R$) and the left ($L$) wedges. Two entangled copies of thermal field theory live separately in the asymptotic boundaries of $R$ or $L$ regions, and they can be non-locally correlated. A viable measure of such correlation can be described by an appropriate generalization of mutual information (MI) \cite{Wolf:2007tdq, Headrick:2007km, Allais:2011ys, Fischler:2012uv, Hayden:2011ag} known as thermo-mutual information (TMI) first introduced in \cite {Morrison:2012iz} along with a genralization of holographic mutual information (HMI) as holographic thermo mutual information (HTMI). One of the contributions to HTMI requires a wormhole like extremal surface that connects the two asymptotic boundaries. MI and TMI share many common properties such as both of them are positive definite and free of UV singularities  etc.
However, at low temperature, HMI shoots to a very high value whereas HTMI vanishes. Moreover, TMI provides a lower bound
on MI itself.

The correlations between two boundary theories can be disrupted by the exponential growth of an early time perturbation introduced in one of the boundary field theories \cite{Leichenauer:2014nxa, Mezei:2016wfz, Mezei:2016zxg, Jahnke:2017iwi, Fischler:2018kwt, Cai:2017ihd, Yang:2018gfq, Sircar:2016old}. Such exponential growth of perturbation can also be used to study the butterfly velocity and the Lyapunov exponent as the diagnosis of chaos \cite{Shenker:2013pqa}.

In this work, we consider that each of these boundary field theories forming the TFD state is backreacted by the presence of a uniform static distribution of heavy fundamental quarks \cite{Chakrabortty:2011sp, Chakrabortty:2020ptb}. We expect that the purification of the boundary theory can still be possible in the form of a backreacted TFD state. 
 The bulk theory dual to the backreacted TFD state is now described as a deformed eternal (two sided) black hole, where each side of that is deformed by a uniform, static distribution of long fundamental strings attached to the asymptotic boundary and stretched along the radial direction up to the horizon. 
In particular, the $(d+1)$-dimensional action for each side of the two-sided geometry is described as,
 \begin{equation}
    \mathcal{S}
=
    \frac{1}{4 \pi G_{N}} \int d x^{d+1} \sqrt{g}(\mathcal{R}-2 \Lambda)+\mathcal{S}_{M},~~~~ \mathcal{S}_{M}=  -\frac{1}{2} 
   \sum_{i=1}^N    \mathcal{T} \int d^{2} \xi \sqrt{-h} h^{\alpha \beta} \partial_{\alpha} X^{\mu}_i \partial_{\beta} X^{\nu}_i g_{\mu \nu} ,
 \end{equation}
 where $\mathcal{T}$ is the tension of a single string, $g_{\mu \nu}(x)$ and $h_{\alpha \beta}(\xi)$ are the bulk metric and the worldsheet metric respectively. There exists an exact analytical solution to the Einstein equation which nicely incorporates the holographic realization of the boundary deformation \cite{Chakrabortty:2011sp},
 \begin{equation}
   d s^{2}
=
    \frac{R_{\text{AdS}}^{2}}{z^{2}}\left(-h(z) d t^{2}+d \vec{x}^{2}+\frac{d z^{2}}{h(z)}\right)
,~~
    h(z)= 1-\frac{2 m}{R_{\text{AdS}}^{2 d-2}} z^{d}-\frac{2 b}{(d-1) R_{\text{AdS}}^{d-1} }z^{d-1},\label{eq:gmunu}
 \end{equation}
where $b = { \mathcal{T} N R_{\text{AdS}}^{d+1}}/{ V_{d-1} }$ is the string backreaction parameter, and $N$ is the number of strings.
The Hawking temperature is given by $
4\pi T=-{dh(z)}/{dz}|_{z=z_H}$,

With this holographic set up, we compute the HTMI that may take a contribution from an extremal surface connecting two asymptotic boundaries of the deformed eternal blackhole. Following \cite {Shenker:2013pqa}, we compute the butterfly velocity and the Lyapunov exponent by introducing a shock wave perturbation in one side of the deformed eternal blackhole. Note that here we consider two types of geometries. One of them is a two-sided AdS blackhole, where each side is deformed in a non-perturbative way by uniform distributions of strings (string backreaction). Further, we introduce shock wave perturbation in the R region of the two-sided string-deformed geometry and solve the Einstein equation perturbatively. The perturbative solution is an example of shock wave geometry in AdS spacetime \cite {Sfetsos:1994xa} where the wormhole region gets stretched and as a consequence of that we observe a degradation of HTMI. We quantify the time variation of such degradation of HTMI in terms of entanglement velocity. We emphasise the role of backreaction parameter ($b$) in the characterization of the chaos and in the estimation of the correlation in this work.

The organization of the paper is as follows. In section \ref{sec:NonperturbativeTFD} we calculate the HTMI in deformed eternal  black hole geometry. In section \ref{sec:Shockandvb}, we perform the shockwave analysis and holographically compute the butterfly velocity and the Lyapunov exponent. In section \ref {sec:ShockandMI}, we study the rate of degradation of HTMI due the presence of shock backreaction and calculate the entanglement velocity. Finally, we summarize our work in section \ref{sec:summary}.

\begin{figure}
\centering
  \includegraphics[width=.8\linewidth]{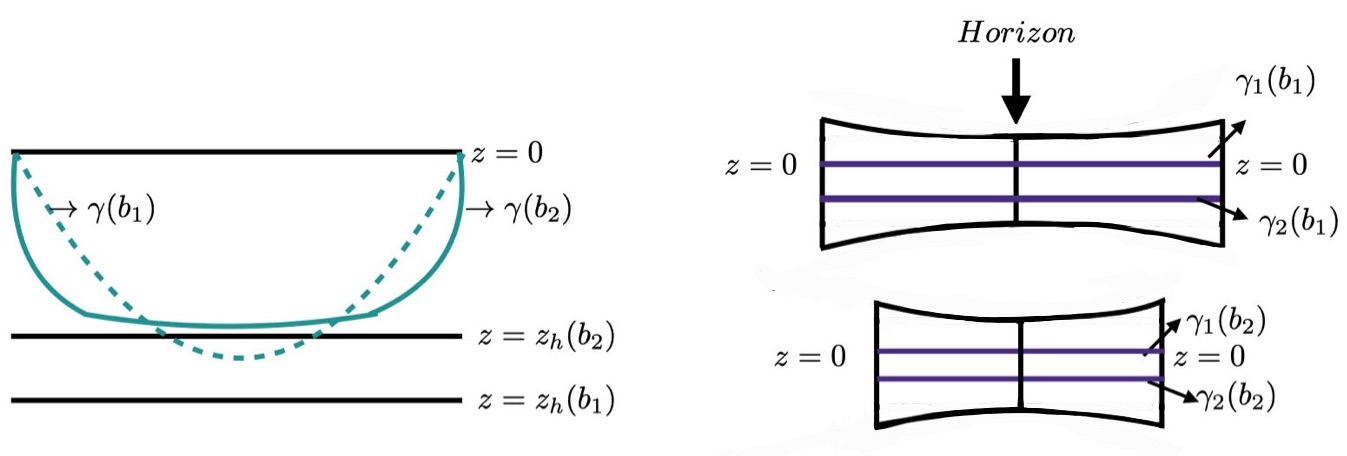}
  \caption{\textbf{Left -} The shape of two RT surfaces, viz. $\gamma(b_1)$ and $\gamma (b_2)$  where $b_1$ and $b_2$ are two distinct values of string cloud densities. Note that if $b_1 < b_2$ then the shapes of $\gamma(b_1)$ and $\gamma (b_2)$ are  adjusted to hold the inequality $\mathcal{A}(\gamma(b_1))< \mathcal{A}(\gamma(b_2)).$ \textbf{Right -} The top diagram corresponds to the exterior of a two-sided blackhole deformed by  $b_1$ and the bottom one corresponds to a deformation made by $b_2$. The length of the bottom one is squeezed as compared with the length of the top one. Consequently, the length of the extremal surface connecting two asymptotic boundaries at $t=0$ corresponding to top one is longer than the same for the bottom one.}
\label{noshockcartoon}
\end{figure}

\section{Effect of backreaction on HTMI in two sided geometry}
\label{sec:NonperturbativeTFD}

In this section, we study the effect of quark backreaction on the TMI between two non-interacting boundary field theories. The TMI between subsystems $A$ and $B$ defined as, $I(A,B)=S(A)+S(B)-S(A \cup B )$
where $S(A)$, $S(B)$ and $S(A \cup B)$ are the entanglement entropies (EEs) of $A$, $B$ and $A \cup B$ respectively.

Following \cite{Leichenauer:2014nxa}, we compute HTMI for strip like identical subsystems $A$ and $B$ of width $l$ on the asymptotic boundaries of $R$ and $L$ regions of deformed eternal blackhole respectively. Note that the HTMI is a linear combination of holographic entanglement entropies (HEEs). According to the Ryu-Takyanagi (RT) prescription \cite{Ryu:2006bv}, the HEE is defined as, $S= {\mathcal{A}(\gamma)}/{4G_N}$, where $\mathcal{A}$ is minimal area of the extremal surface $\gamma$ homologous to entangling region on the boundary. 
In particular,  $S(A)$ and $S(B)$ are proportional to the minimal areas of the extremal surfaces $\gamma_A$ and $\gamma_B$ in the bulk, homologous to entangling regions $A$ and $B$ respectively, where $\gamma_i$ is defined by the embedding ($t=0, z, -l/2 \leq x (z) \leq l/2, -L/2 \leq x^j \leq L/2, i = A, B $ and $j=2, \cdots, d-1$).
The candidates for the extremal surface $\gamma_{A\cup B}$ can be either $\gamma_A \cup \gamma_B$ or $ \gamma_1 \cup \gamma_2 = \gamma_{\text{wormhole}}$ where $\gamma_1$ and $\gamma_2$ correspond to the embeddings ($t=0, z, x=-l/2, -L/2 \leq x^j \leq L/2$) and ($t=0, z,  x=l/2, -L/2 \leq x^j \leq L/2$) respectively, connecting two asymptotic boundaries  through the bifurcation point of the deformed eternal blackhole.

\begin{figure}[h]
\centering
  \includegraphics[width=.40\linewidth]{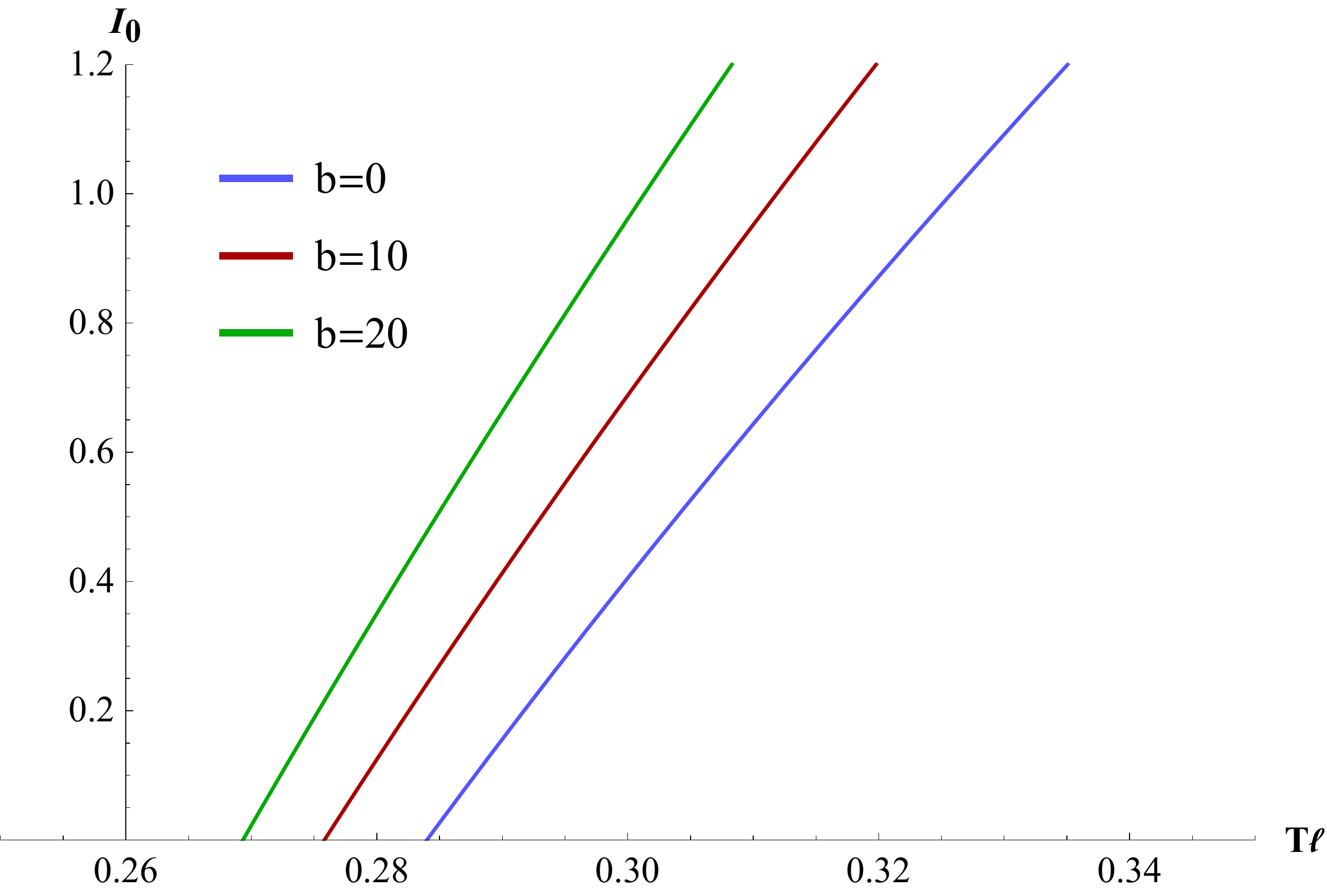}
  \hspace{.15\linewidth}
  \includegraphics[width=.40\linewidth]{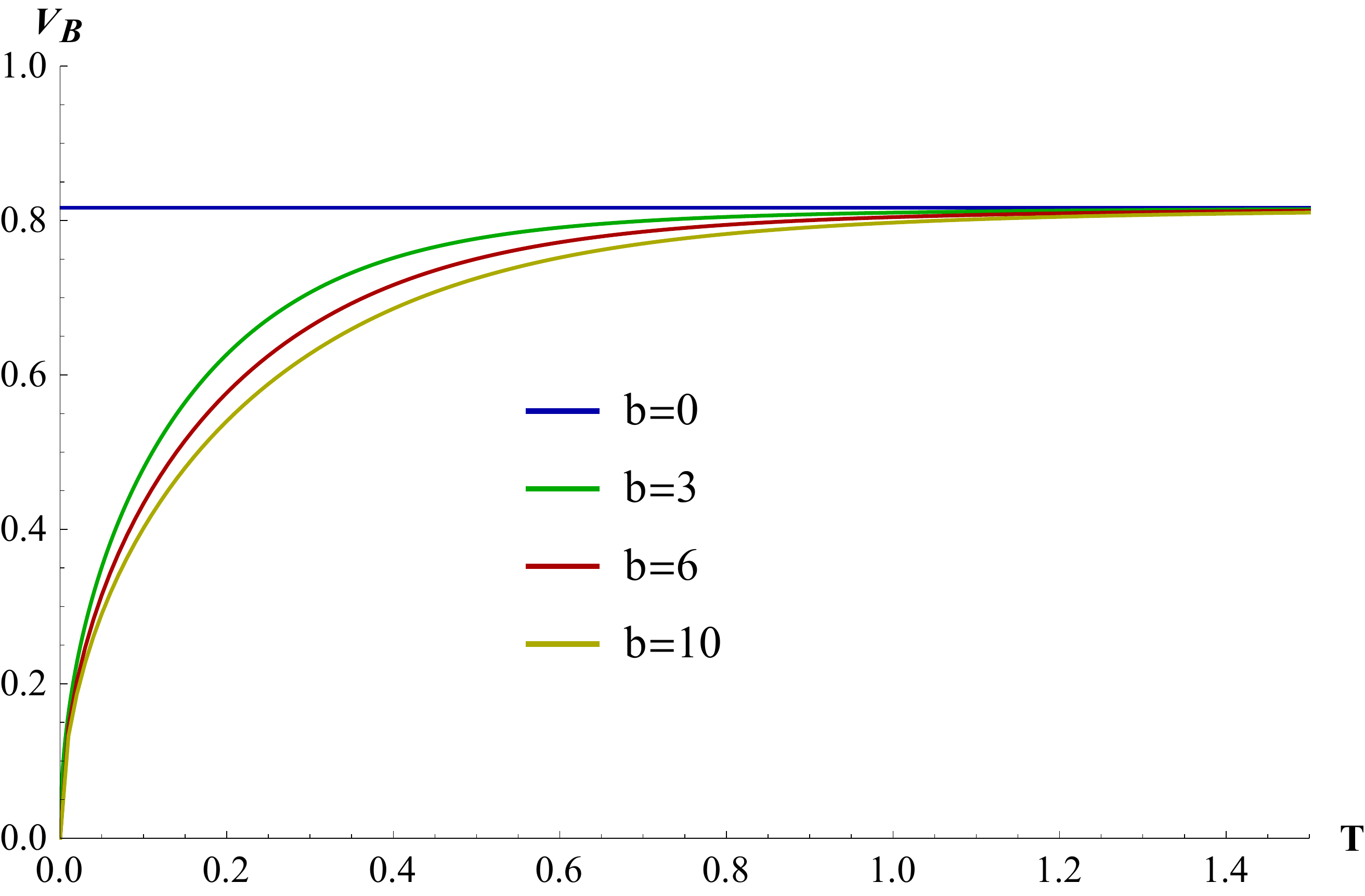}
  \caption{\textbf{Left-} TMI is plotted with respect to the width $l$ at constant temperature $T = 1$ for $d=4$ and $b=0,10,20$.
  \textbf{Right-} The butterfly velocity $v_B$ in d=4 is plotted with respect to the temperature $T$ for four distinct values of $b$.  The plot shows that as $T$ increases $v_B$ takes higher values. For a fixed temperature, $v_B$ increases as $b$ takes lower values.}
\label{Ivslwithoutshock}
\end{figure}

If $\mathcal{A}(\gamma_A \cup \gamma_B)$ is less than $\mathcal{A}(\gamma_{\text{wormhole}})$ then $I (A,B)$ is zero, and conversely $I(A,B)$ is positive in the opposite situation. When $\gamma_{A\cup B} = \gamma_{\text{wormhole}}$, the computation of $S(A\cup B)$ follows from the Hubeny-Rangamani-Takayanagi (HRT) prescription \cite {Hubeny:2007xt} where the determinants of induced metrics corresponding to the extremal surfaces are, 
\begin{equation}
    G_{in}^A= G_{in}^B=\left(R_{\text{AdS}}^2/z^2\right)^{d-1}\left(1/h(z)+x'^2\right)
,
\qquad
    G_{in}^{wormhole}
=
    \left(
    R_{\text{AdS}}^2/z^2
    \right)^{d-1}
  /h(z)
,
\end{equation}

where $x'=dx/dz$, and it is zero for $\gamma_1$ and $\gamma_2$. Now, the holographic prescription suggests the following form of HTMI,
\begin{equation}
\begin{split}
    I(A,B)
&= 
   \frac{1}{4 G_N} 
   \left(
   \prod _{i=2}^{d-2} 
    \int_{-L/2}^{L/2} 
    dx^i
    \right)
    \left[  
    2
    \int_{0} ^{z_t}
    dz 
    \left(
    \sqrt{ G_{in}^{A} } 
    +
    \sqrt{G_{in}^{B}}
    \right)
    -  
    4
    \int_{0} ^{z_H}
    dz
    \sqrt{G_{in}^{A\cup B}}
    \right]
\\   
&=
    \frac{
    L^{d-2}R_{\text{AdS}}^{d-1}}{G_N} 
    \left[
    \int_{0}^{z_t}\frac{dz}{z^{d-1}\sqrt{h(z)}\sqrt{1-\left(\frac{z}{z_t} \right)^{2d-2}}}-\int_{0}^{z_H}\frac{dz}{z^{d-1}\sqrt{h(z)}}
    \right].
\label{eq:mi}
\end{split}
\end{equation}
The factor $2$ and $4$ appearing in front of the integrations in the first line of \ref{eq:mi} are the consequences of the symmetric structure of the extremal surfaces. $z_t$ is the turning point of the extremal surfaces (RT surfaces) corresponds to region $A$ and $B$. The variation of the mutual information with respect to the width $l$ of the entangling region $A$ or $B$ can be obtained by using the  relation between $l$ and $z_t$,
\begin{equation}
l/2=1/z_t^{d-1}\int_{0}^{z_t} z^{d-1}\left[h(z)\left(1-(z/z_t)^{2d-2}\right)\right]^{-1/2} dz
\end{equation} 
There exists a critical width $l_{c}$ below which the HTMI becomes zero. Throughout this work we consider $l\geq l_c$. The behaviour of HTMI with respect to the width $l$ for a constant  temperature $T=1$ is shown in the figure \ref{Ivslwithoutshock} (left), which can be summarized as follows.
\begin{itemize}
    
    \item It is evident from the plot that in $d=4$, for a given temperature, the mutual information increases with respect to quark cloud density. Note that it has been observed in \cite{Chakrabortty:2020ptb} that the HEE's $S(A)$ and $S(B)$ increases with $b$. Since $S(A)$ and $S(B)$ is proportional to $\mathcal{A}(\gamma_A)$ and $\mathcal{A}(\gamma_B)$ respectively, as explained in the figure \ref{noshockcartoon} (left), this is possible only when $x'$ decreases with higher values of $b$. Also according to \ref{noshockcartoon} (right) the two sided HEE $S(A\cup B)$ decreases with $b$. Therefore these two facts combined together explain the enhancement of HTMI with respect to $b$ shown in figure \ref{Ivslwithoutshock} (left). This is also  expected to be true in higher dimensions.  
     \item This critical width $l_c$  decreases as we increase the quark density parameter $b$. 
    
    \item  For any value of width $l\leq l_c$ TMI between two subsystems will be identically zero which indicates the absence of any type of correlations between two subsystems $A$ and $B$.
      
\end{itemize}

\section{Shock wave analysis: Butterfly velocity and Lyapunov exponent }
\label{sec:Shockandvb}

In this section, we characterize the chaos in  large-$N_c$ thermal field theory backreacted due to the presence of a uniform distribution of heavy quarks, by evaluating the butterfly velocity and Lyapunov exponent. In order to do that we consider a tiny in-going perturbation in the form of a  pulse of energy $E_0$, thrown in to the $R$ region of the string backeracted eternal blackhole, from the asymptotic boundary at early time. As the pulse evolves in time and falls into the horizon at a late time corresponding to the boundary time $t=0$, it gets blue-shifted, becomes a shockwave and deforms the bulk spacetime to a shock wave geometry.

Without shockwave the two-sided string deformed black hole geometry has the following metric written in the Kruskal-Szekeres coordinate $(U,V)$ as
\begin{equation}\label{unper}
    ds^2
= 
    2A(U,V)dUdV+g_{xx}(U,V)d\vec{x}_{d-1}^2
,
\qquad
    A(U,V)
=
    \frac{\beta^2}{8\pi^2} 
    \frac{
    \left| g_{tt}(U,V) \right|
    }{
    UV
    }
,
\end{equation}
where $
    U
= 
    \sigma_U
    \exp 
    \left( 
    - \frac{2 \pi}{\beta} \left( t - z_*\right)
    \right)
,~
    V
= 
    \sigma_V
    \exp 
    \left( 
    \frac{2 \pi}{\beta} \left( t + z_*\right)
    \right)
$, and $
    z_*
=
    -
    \int_0^z
    \sqrt{
    1/h(z)^2
    }
    dz    
$.
The coefficients $\sigma_U$ and $\sigma_V$ take values $\pm 1$ depending on the region of the two sided blackhole spacetime we consider. For example, $\sigma_U=-1$ and $\sigma_V=1$ corresponds to the right exterior region $R$ of the Penrose diagram in figure \ref{pen}. The most general form of energy momentum tensor that corresponds to the metric in (\ref{unper}) is given by, 
\begin{equation}
\begin{split}
    T^{\text{matter}}_0
=&
    2
    T_{{U}{V}} 
    dU
    dV
    +
    T_{{U}{U}}d{U}^2
    +
    T_{{V}{V}}d{V}^2
    +
    T_{ij}d{x^i}d{x^j}
,\qquad T_{\mu \nu} = T_{\mu \nu} (U,  V, \vec{x}).
\label{eq:Tmatteroriginal}
\end{split}
\end{equation}

The shockwave geometry is described by the shifted Kruskal-Szekeres coordinates as follows,
\begin{equation}
\begin{split}
    U
&\to 
    U+\theta(V) \alpha(t,\vec{x})
,\qquad    
    V
\to 
    V
,
\\
    dU 
&\to 
    dU+\theta(V) d\alpha(t,\vec{x})
,\qquad    
    dV
\to 
    dV
.
\label{eq:shifts}
\end{split}
\end{equation}
Here the theta function represents the fact that only the region with $V>0$ is modified due to the shockwave parameter $\alpha$. 

We can redefine the Kruskal-Szekeres coordinates as  $
    \hat{U}
=
    U+\theta(V) \alpha(t,\vec{x})
,\hat{V}
=
    V
, \text{and}~\hat{x}^i=x^i$.
The shockwave geometry is realized by being substituted the shift (\ref{eq:shifts}) into (\ref{unper}),
\begin{equation}
    ds^2
=
    2
    {A}(\hat{U},\hat{V})d\hat{U}d\hat{V} 
    -
    2
    {A}(\hat{U},\hat{V})
    {\alpha}
    \delta(\hat{V})d\hat{V}^2
    +
    g_{xx}(\hat{U},\hat{V})
    d\vec{\hat{x}}^2
\label{pm}
\end{equation}
with the total energy momentum tensor $T^{\text{matter}}+T^{\text{shock}}$.
The matter part is modified as,
\begin{equation}
\begin{split}
    T^{\text{matter}}
=&
    2
    \left[
    T_{{U}{V}} 
    -
    T_{{U}{U}}
    \alpha\delta(\hat{V})
    \right]
    d\hat{U}
    d\hat{V}
    +
    T_{{U}{U}}d\hat{U}^2
    +
    T_{ij}d{x^i}d{x^j}
\\
    &+
    \left[
    T_{{V}{V}}
    -
    2
    T_{{U}{V}}
    \alpha
    \delta(\hat{V})
    +
    T_{{U}{U}}
{\alpha}^2 
    \delta(\hat{V})^2
    \right]
    d\hat{V}^2 
,
\end{split}
\end{equation}

where $T_{\mu \nu} = T_{\mu \nu} (\hat U, \hat V, \vec{x}) $.
The energy momentum tensor for the shockwave is given by
\begin{equation}
    T^{\text{shock}}
=
    E_0
    \exp{
    \left(\frac{2\pi }{\beta}t\right)
    }
    \delta(\hat{V}) 
    \delta (x^i)
    d\hat{V^2}
,
\end{equation}
where $E_0$ is a constant related to asymptotic energy of the pulse, $\exp{
    \left(\frac{2\pi }{\beta}t\right)}$ is the blue shift factor and $\delta(x^i)$ represents the fact that the source is localized at $x^i = 0$. To obtain the shift function $ \alpha$, we solve the following Einstein's equation, 
\begin{equation}
    \mathcal{R}_{\mu \nu}
    -
    \frac{1}{2} 
    g_{\mu \nu}
    \mathcal{R}
    +
    \Lambda
    g_{\mu\nu}
=
    8 \pi G_{N}
    \left(
    T_{\mu \nu}^{\text {matter }}
    +
    T_{\mu \nu}^{\text {shock }}
    \right)
.
\end{equation}
To keep track of the order of perturbation we multiply the source and the effect of geometric shift with a constant book keeping parameter $\epsilon$: $T_{\mu \nu}^{\text {shock }} \to \epsilon T_{\mu \nu}^{\text {shock }}$ and $ \alpha \to \epsilon  \alpha$, where $\epsilon \ll 1$.

Let the shift function satisfy $ \partial \alpha / \partial x^{i} \not= 0$, i.e., the perturbation propagates in the $x^i$-direction.
In addition to this assumption, our metric (\ref{eq:gmunu}) in terms of the modified Kruskal-Szekeres coordinates satisfies $\partial A (\hat U, \hat V)/\partial {\hat U} = 0$ and $\partial g_{ij}(\hat U, \hat V)/\partial {\hat U} = 0$ at $\hat V=0$.
With these conditions, we obtain $T^{\text{matter}}_{UU}|_{\hat V=0}=0$ from the zeroth order of Einstein equation in $\epsilon$.  
Finally, the $\hat U \hat V$  component of Einstein equation in  the first order of $\epsilon$ gives,
\begin{equation}
g^{i j} \left(z_H \right) \left[ A\left(z_H\right) \partial_{i} \partial_{j} - \frac{z_H}{2} \partial_z g_{i j} \left(z_H\right) \right]  \alpha\left(t, \vec{x}\right)
= 8 \pi G_{N} E_0 \exp \left(\frac{2 \pi t}{\beta}\right)\delta\left(x^i\right),
\label{eq:VVcomp}
\end{equation}
now in terms of the coordinates $(t,z,\vec{x})$.

Note that $\Lambda$ does not appear in (\ref{eq:VVcomp}), since that quantity is canceled out by counterparts, which are obtained from relations of the zeroth order of Einstein equation in $\epsilon$. Simplifying (\ref{eq:VVcomp}), we obtain the following differential equation for $\alpha(t,x)$,
\begin{equation}
    \left(
    \sum_{i=1}^{d-1}
    \partial_i^2
    -
    M^2
    \right)
    \alpha(t,\vec{x})
=
    \frac{g_{xx}(z_H)}{A(z_H)} 
    8\pi E_0
    \exp{\left(\frac{2\pi}{\beta}t\right)}\delta(x^i)
,
\qquad
    M^2
=
    \frac{(d-1)z_H}{2A(z_H)}\partial_{z}g_{xx}(z_H)    
.
\label{dfeoc}
\end{equation}

For large $\absN{x^i}$,  the solution of (\ref{dfeoc}) is obtained as 
\begin{equation}
    \alpha(t,\vec{x})
 \sim 
    \exp{\left[\frac{2\pi}{\beta}(t-t_*)-Mx^i\right]}.
    \label{eq:solalpha}
\end{equation}
In quantum many-body system, the butterfly effect is characterized by the following double commutator involving two local hermitian operators,
  $C(t, x)=-\angleN{ [ W(x, t), V(0)]^{2} }_{\beta}$. The quantity $C(x,t)$ captures how an early perturbation $V(0)$ changes some later measurement $W(x,t)$. It is known that the holographic approach for large-$N_c$ theories gives $C(t, x) \approx \frac{K}{N_c^{2}} \exp\squareN{ \lambda_{L} (t-t_{*}-x / v_{\mathrm{B}}) }$ for $t \lesssim t_*$ \cite{Shenker:2013pqa,Shenker:2014cwa,Roberts:2014isa,Shenker:2013yza}, where $t_*$ is the scrambling time, $\lambda_L$ is the Lyapunov exponent, and $v_B$ is the butterfly velocity. 

Comparing (\ref{eq:solalpha}) to the correlator $C(t, x)$, we obtain 
\begin{equation}
    \lambda_L
= \frac{2\pi}{\beta},
\qquad
{v}_B = \frac{2\pi}{\beta M}
= \sqrt{\frac{d-\rho}{2(d-1)}},
\end{equation}
where $\rho = \frac{2b}{d-1} \roundLR{ \frac{z_H}{R_{\text{AdS}}} }^{d-1}$, and the value of $z_H$ (and $\rho$) is given by two independent parameters such as $T$ and $b$.
Note that $\rho$ is proportional to $b$, and it vanishes when $b=0$.

\begin{figure}
\centering
 \includegraphics[width=.43\linewidth]{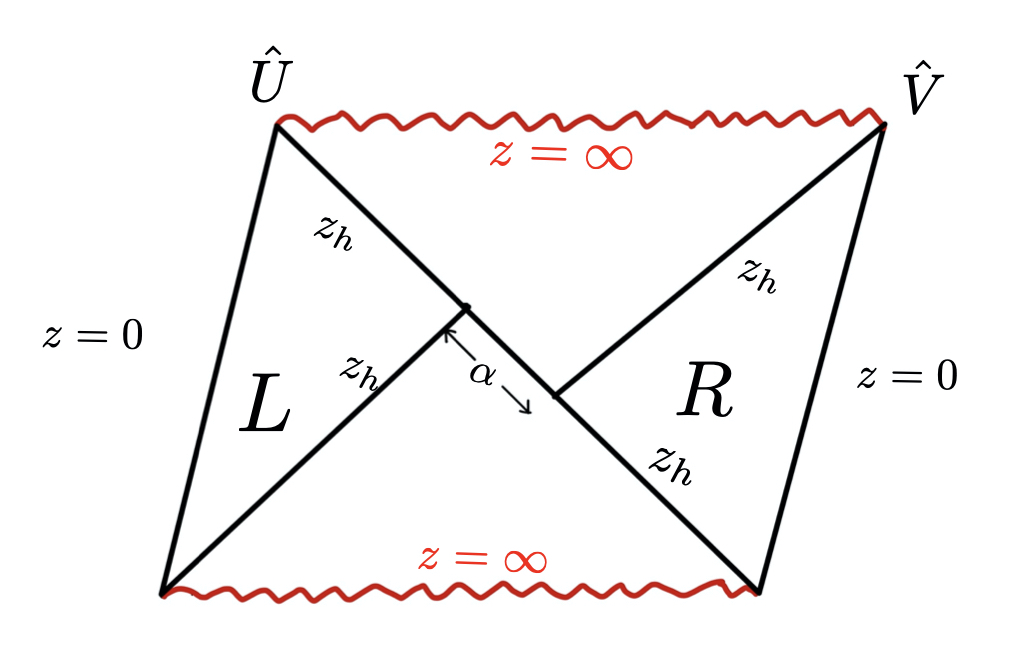}
 \hspace{.10\linewidth}
 \includegraphics[width=.40\linewidth]{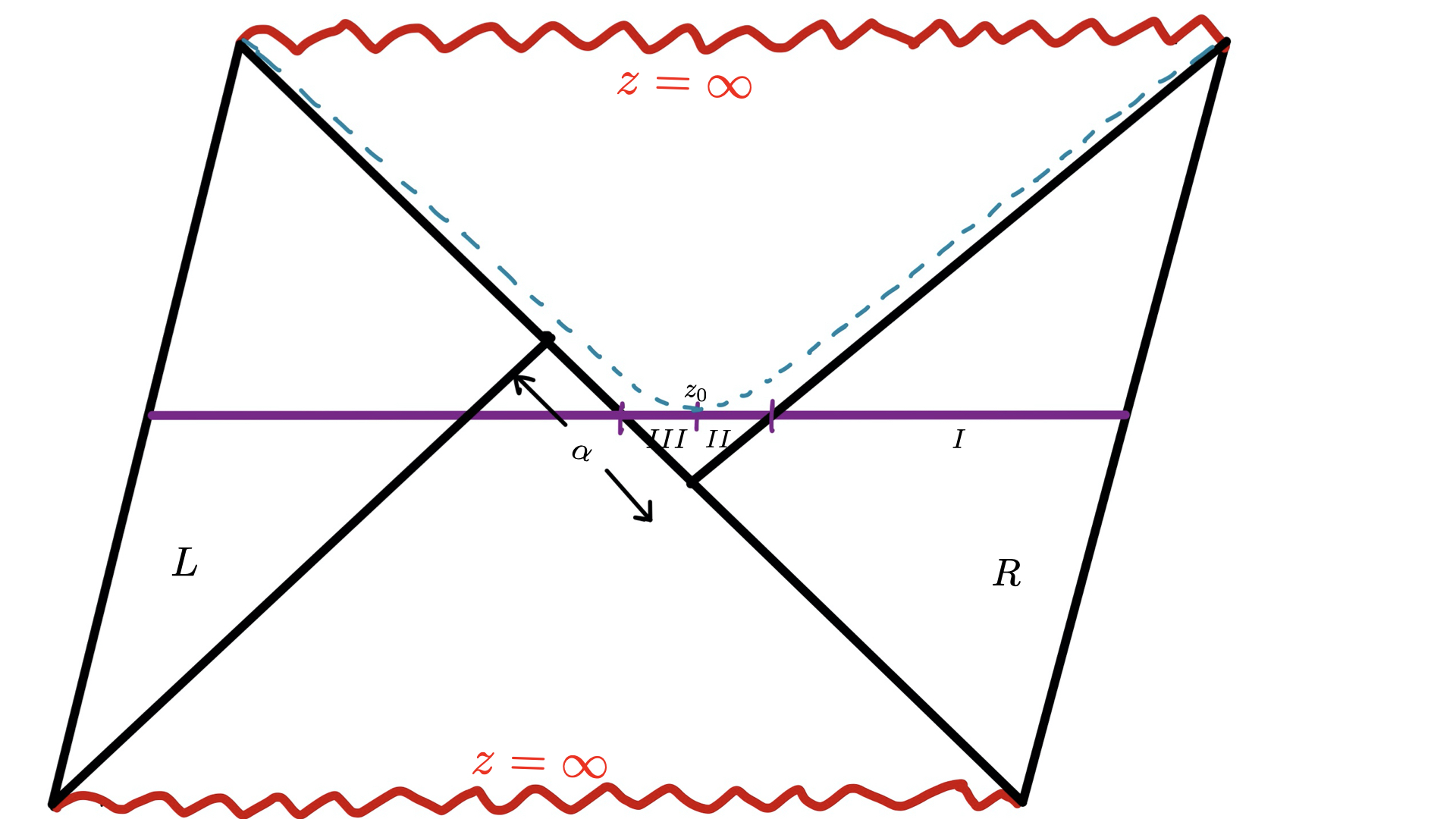}
\caption{\textbf{Left-} Penrose diagram of the deformed eternal blackhole in presence of shock.
\textbf{Right-} The horizontal violet line is the extremal surface in the shock wave geometry. We 
consider three regions in the right half of the geometry, I, II and III. Note that $z_0$ is the intersection point between extremal curve and the constant $z_0$ line (dashed, blue) that determines the fact that  II and III have the same area.}
\label{onlyshockpen}
\end{figure}

For $d=4$ and $R_{\text{AdS}}=1$, we plot the $T$ dependence of $v_B$ for several values of $b$, in figure \ref{Ivslwithoutshock} (right).
\begin{itemize}
\item
For $b=0$, the butterfly velocity attains a constant value $v_B=\sqrt{d/2(d-1)} \approx 0.82$ and remains unchanged as $T$ is varied. 

\item
For finite $b$, the behavior of $v_B$ with respect to $T$ is quite different from the $b=0$ case. In particular, in the low temperature regime, $v_b$  starts from zero for all values of $b$ and increase within the intermediate values of temperature. Finally $v_B$ saturates in the high temperate reion, to the value $v_B=\sqrt{d/2(d-1)}$ for all $b$. In other words, the exponential growth of the perturbation characterizing the chaos expands with $v_B$ only in the intermediate region of temperature for any nonzero $b$.


\end{itemize}

\section{\textbf{Disruption of Thermo Mutual Information due to scrambling}}
\label{sec:ShockandMI}

In this section, we perform a holographic analysis of HTMI in the shockwave geometry. The spatially homogeneous shock parameter $\alpha$ we consider here is a special case of what we have discussed in the previous section.


The HTMI in the presence of shock perturbation can be defined as,
\begin{equation}
\label{MIreg}
 I(A,B;\alpha)= I(A,B;\alpha=0)- S^{reg}(A\cup B;\alpha),
\end{equation}
where $I(A,B;\alpha=0)$ is already computed in equation (\ref{eq:mi}) and to avoid the $\alpha$ independent UV divergences we define a regularized version of HEE, $S^{reg}(A\cup B;\alpha)=S(A\cup B;\alpha)-S(A\cup B;\alpha=0)$.
 In order to compute $S^{reg}(A\cup B;\alpha)$ we choose a set of appropriate time dependent embeddings defined as $\{t, z(t),  x=-l/2, -L/2 \leq x^j \leq L/2\}$ and $\{t, z(t),  x=l/2, -L/2 \leq x^j \leq L/2\} $ (figure \ref{onlyshockpen} (right)). The area functional corresponding to either of these time dependent embeddings is given as,
\begin{equation}
\mathcal{A}=L^{d-2}R_{AdS}^{d-1} \int{{\frac{dt}{z^{d-1}} {\Big(-h(z)+\frac{\Dot{z}^2}{h(z)}\Big)}^{1/2}}},\qquad \mathcal{L}= {\frac{L^{d-2}R_{AdS}^{d-1}}{z^{d-1}} {\Big(-h(z)+\frac{\Dot{z}^2}{h(z)}\Big)}^{1/2}} 
\end{equation}

 The Lagrangian density does not explicitly depend on time and that allows us to find the corresponding conserved quantity $\mathcal{C}$ with the condition $\Dot{z}=0|_{z=z_0}$,~ $\mathcal{C}=-\left(\frac{R_\text{AdS}}{z_{0}}\right)^{d-1}\sqrt{-h(z_{0})}$. Now, the onshell action and the  time coordinate along the minimal surface can be expressed as,
\begin{equation}
\mathcal{A}=L^{d-2}R_\text{AdS}^{d-1} \int{{\frac{dz}{z^{d-1}} \frac{1}{\sqrt{ h(z)+\mathcal{C}^2z^{2d-2}}}}}, \qquad t(z)=\sigma \int{\frac{dz}{h(z)\sqrt{1+\mathcal{C}^{-2} h(z)z^{2-2d}}}}.
\label{wharea}
\end{equation}


To find $\mathcal{A}(\gamma_{\text{wormhole}})$, we break the domian of integration variable $z$ given in equation (\ref{wharea}) into three regions. The first region (I) starts from the right boundary  and goes into the bulk up to the horizon. The second region (II) starts from the horizon and ends at  $z=z_0$. The third region (III) starts from $z=z_0$ goes to the left (see the figure \ref{onlyshockpen} (right)) and extends up to horizon.  
 Here, $\sigma$ takes the value $\pm 1$ depending on the rate of change of z with respect to $t$ along the extremal surface.

Considering the three regions I, II and III, now the explicit form of $\mathcal{A}(\gamma_{\text{wormhole}})$ becomes, 

\begin{equation}
\mathcal{A}(\gamma_{\text{wormhole}}) = 
4 L^{d-2}R_\text{AdS}^{d-1} \left[\int^{z_H}_{0}{{\frac{dz}{z^{d-1}}\biggl( \frac{1}{\sqrt{ h(z)+\mathcal{C}^2z^{2d-2}}}-\frac{1}{\sqrt{ h(z)}}}}\biggr)+2 \int^{z_0}_{z_H}{{\frac{dz}{z^{d-1}} \frac{1}{\sqrt{ h(z)+\mathcal{C}^2z^{2d-2}}}}}\right],
\end{equation}
and the corresponding regularized HEE is 
$S^{reg}_{A\cup B}(z_0) = \mathcal{A}(\gamma_{\text{wormhole}}) / 4 G_N$.

\begin{figure}
\centering
 \includegraphics[width=.40\linewidth]{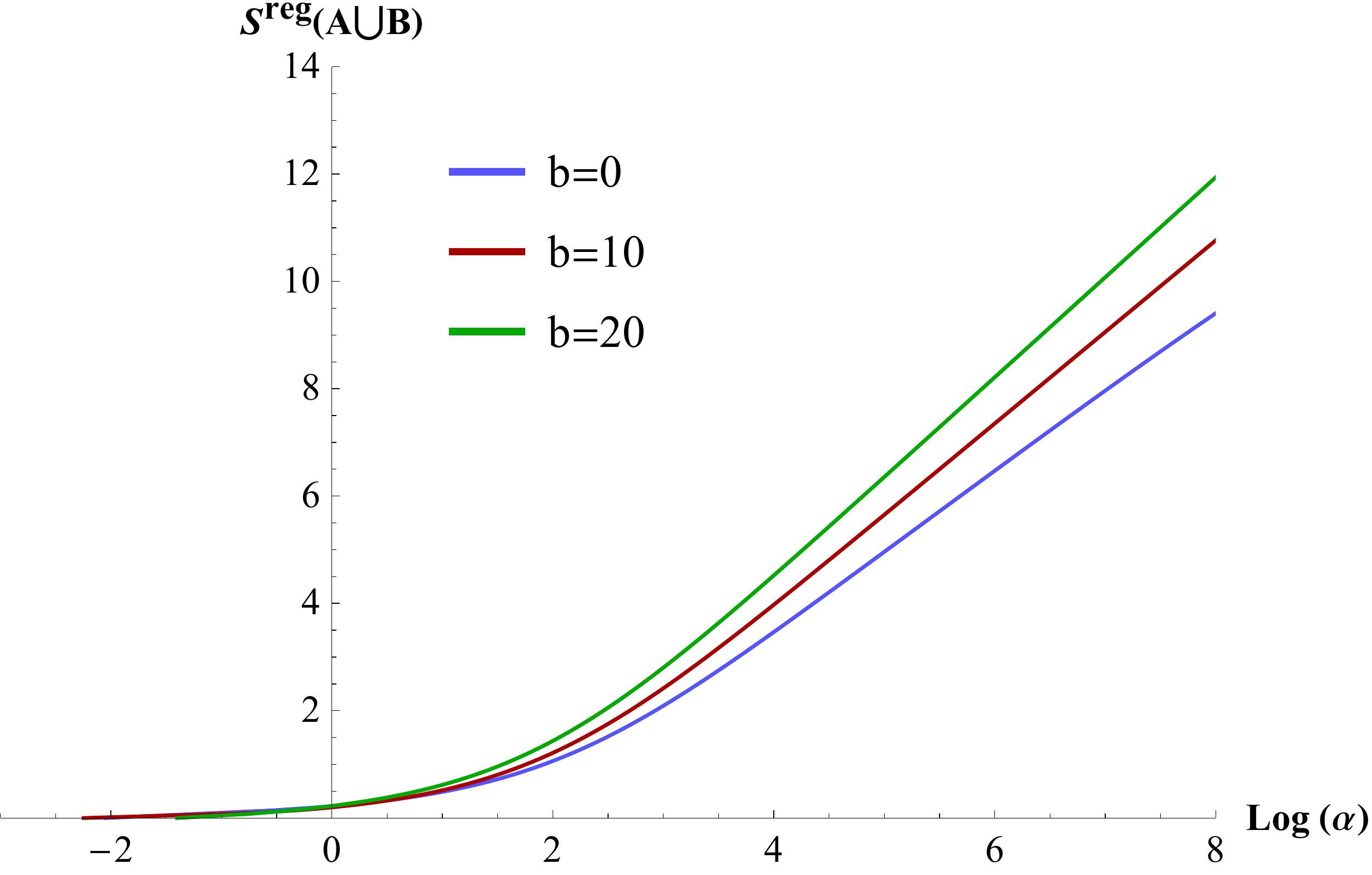}
 \hspace{.10\linewidth}
 \includegraphics[width=.40\linewidth]{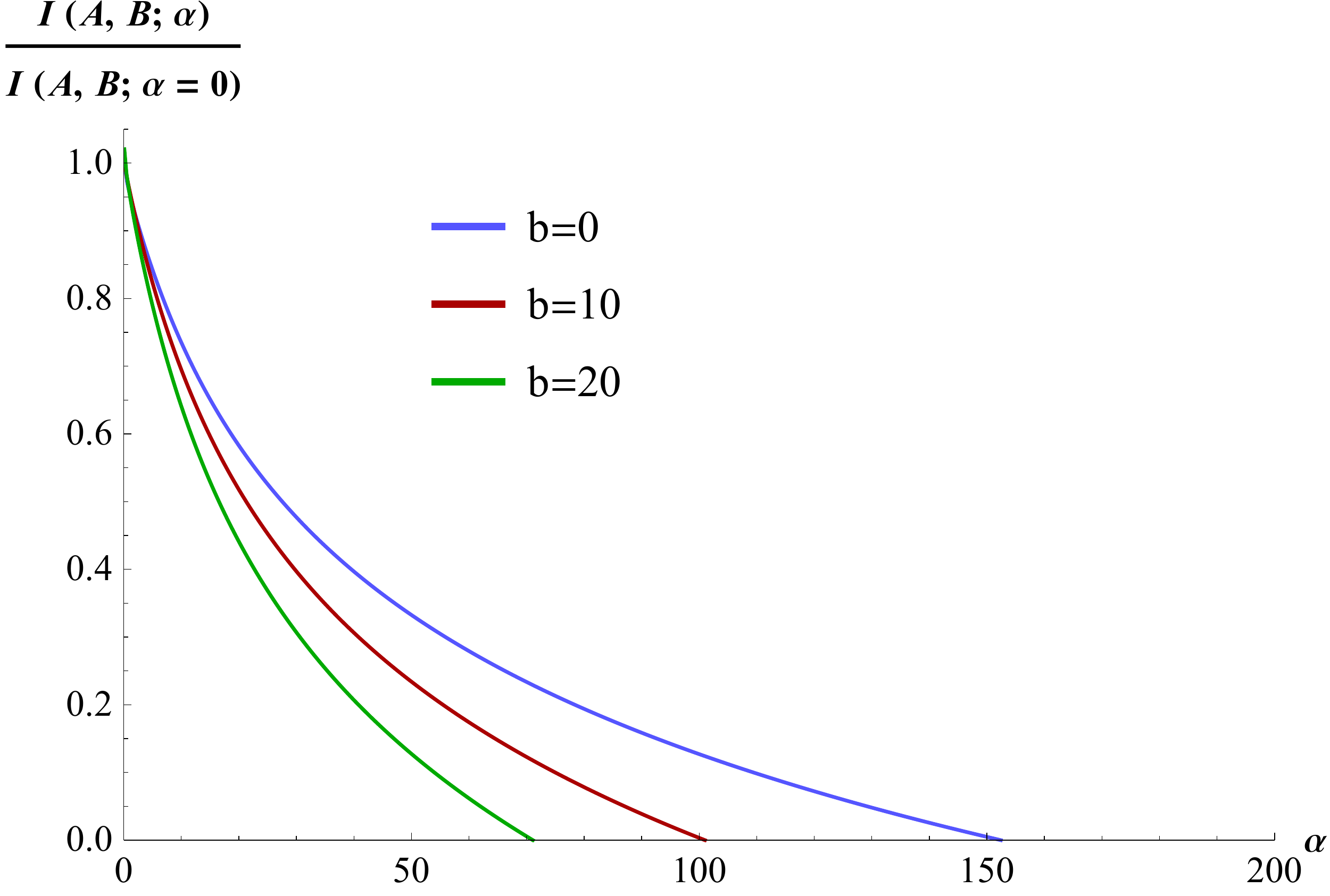}
  \caption{\textbf{Left-} Plot for $S^{reg}(A\cup B)$ with respect to $\log(\alpha)$ for $d=4$ and different values of $b$. This plot shows that regularized entropy increases as string cloud density increases for fix $\alpha$.
  \textbf{Right-} Plot for TMI in presence of shock wave for $d=4$ with respect to shock wave parameter $\alpha$ for different values of $b$. This plot shows that TMI starts from a fixed value and gradually decreases and vanishes for a particular value of $\alpha$.}
\label{Sreg}
\end{figure}

However, in order to see the variation of $S^{reg}_{A\cup B}$ with respect to the shock wave parameter, we need to find the relation between $z_0$ and $\alpha$. As described in figure \ref{onlyshockpen} (right), the region I is defined from boundary point $(\hat{U},\hat{V})=(-1,1)$  to  point on horizon $(\hat{U},\hat{V})=(0,\hat{V}_1)$.
Region II is located from $(\hat{U},\hat{V})=(0,\hat{V}_1)$  to $z_0$  $(\hat{U},\hat{V})=(\hat{U}_2,\hat{V}_2)$. Region III is defined from $(\hat{U},\hat{V})=(\hat{U}_2,\hat{V}_2)$ to $(\hat{U},\hat{V})=(\alpha/2,0)$.
Using the definition of the Kruskal coordinates as given in section \ref{sec:Shockandvb}, one can write down the variation of $\hat{V}$ as, 
\begin{equation}
\Delta \log \hat{V}^2=\log \hat{V}_1^2-\log{\hat{V}_0^2} =\frac{4\pi}{\beta}(\Delta z_*+\Delta t),~~  \hat{V}_1 =\exp{\biggl[\frac{2\pi}{\beta}\int^{z_H}_{0}{\frac{dz}{h(z)}\biggl({\frac{1}{\sqrt{1+\mathcal{C}^{-2} h(z)z^{2-2d}}}}-1}\biggr)}\biggr]
\end{equation}

Note that in region I, $\dot{z}>0$ implies $\sigma=+1$ in the expression of $t$. In region II, the negative numerical value of  $h(z)$ corresponds to   $\dot{z}<0$ and thus we set $\sigma=+1$. Now the variation of $\hat{V}$ in region II is the following,
\begin{equation}
\Delta \log\hat{V}^2=\log \hat{V}_2^2-\log{\hat{V}_1^2}=\frac{4\pi}{\beta}(\Delta z_*+\Delta t),~~\hat{V}_2 =\hat{V}_1\exp{\biggl[\frac{2\pi}{\beta}\int^{z_0}_{z_H}{\frac{dz}{h(z)}\biggl(\frac{1}{\sqrt{1+\mathcal{C}^{-2} h(z)z^{2-2d}}}-1\biggr)}\biggr]}
\end{equation}

To compute $\hat{U}_2$ we will consider a reference surface at $\Bar{z}$ inside the horizon such that $z_*=0|_{z=\Bar{z}}$.

\begin{equation}
\hat{U}_2 =\frac{1}{\hat{V}_2}\exp{\biggl[-\frac{4\pi}{\beta}\int^{z_0}_{\Bar{z}}{\frac{dz}{h(z)}}\biggr]}
\label{u2}
\end{equation}
In region III, $\dot{z}>0$ but $h(z)$ still takes the negative numerical value, so we set $\sigma=-1$. Now the variation in $\hat{U}$ coordinate in region III takes the form,
\begin{equation}
\alpha =2\hat{U}_2\exp{\biggl[\frac{2\pi}{\beta}\int_{z_H}^{z_0}{\frac{dz}{h(z)}\biggl(1-\frac{1}{\sqrt{1+\mathcal{C}^{-2} h(z)z^{2-2d}}}}\biggr)\biggr]}
\label{alpha}
\end{equation}
By combining (\ref{u2}) and (\ref{alpha}) we achieve the required relation between $\alpha$ and $z_0$ as follows,
\begin{equation}
\alpha(z_0) =2\exp{[ \mathcal{\eta}_{\text{I}}+\mathcal{\eta}_{\text{II}}+\mathcal{\eta}_{\text{III}}]} ,
\label{az0}
\end{equation}

where,
\begin{equation}\begin{split}
\mathcal{\eta}_{\text{I}} &=\frac{4\pi}{\beta}\int_{z_0}^{\Bar{z}}{\frac{dz}{h(z)}}
,~~~~\mathcal{\eta}_{\text{II}}=\frac{2\pi}{\beta}\int_{0}^{z_H}{\frac{dz}{h(z)}\left(1-\frac{1}{\sqrt{1+\mathcal{C}^{-2} h(z)z^{2-2d}}}\right)}, \\
\mathcal{\eta}_{\text{III}}&=\frac{4\pi}{\beta}\int_{z_H}^{z_0}{\frac{dz}{h(z)}\left(1-{\frac{1}{\sqrt{1+\mathcal{C}^{-2} h(z)z^{2-2d}}}}\right)}
.
\label{k123}
\end{split}
\end{equation}

 With the help of (\ref{az0}) and (\ref{k123}) we study the variations of $S^{reg}(A\cup B)$ and normalized HTMI with respect to the shock wave parameter $\alpha$ for a set of values of $b$ and numerically plot them in figure \ref{Sreg} (left) and in \ref{Sreg} (right). In this analysis we observe the following,
  \begin{itemize}
  
\item In figure \ref{Sreg} (left), we note that for a particular $b$, $S^{reg}(A\cup B)$ increases with $\alpha$.  This behaviour is expected because as we increase $\alpha$, the horizon in the right half of the Penrose diagram in figure \ref{onlyshockpen} (right), shifts towards the boundary. Therefore the intersection point $z_0$ shifts closer to the future singularity and the length of the extremal surface passing through the wormhole   also increases.

\item In figure \ref{Sreg} (right), we find that the normalized HTMI decreases with respect to $\alpha$ when $b$ is fixed. One can also see that more the value of $b$, lower the value of $\alpha$ at which the HTMI becomes zero.

     \end{itemize}

\subsection{Spreading of Entanglement}

In the last section, we have observed a late time growth in $S^{reg}(A\cup B)$ with respect to $\alpha$. In this section, we characterize the growth rate of $S^{reg}(A\cup B)$ in terms of the entanglement velocity $v_{E}$.  In order to compute this growth rate, we first expand $S^{reg}$ about $z_0$ up to linear order and evaluate the same at $z_{0}=z_{c}$ where $\alpha$ diverges.

\begin{equation}
S^{reg}\approx \frac{L^{d-2}R_\text{AdS}^{d-1}\sqrt{-h(z_c)}}{G_N z_c^{d-1}}\frac{\beta}{4\pi}\log{\alpha}, \qquad \text{for}~~ z_0\approx z_c.
\label{srg}
\end{equation}
 In \ref{srg}, since $S^{reg}$ is directly proportional to $\log{\alpha}$, where $\alpha\sim e^{2\pi t_{0}/\beta}$, one can immediately observe the linear growth of the $S^{reg}$ with respect to the boundary time $t_{0}$. The rate of this linear growth is given by,
\begin{equation}\label{a}
\frac{dS^{reg}}{dt_0}= \frac{2L^{d-2}}{R_\text{AdS}^{d-1}}s\biggl(\frac{z_h^{d-1}}{z_c^{d-1}}\sqrt{-h(z_c)}\biggr)=\frac{2L^{d-2}}{R_\text{AdS}^{d-1}}s v_E.
\end{equation}

Here, the thermal entropy density $s$ is defined as,
$s=\frac{R_\text{AdS}^{2d-2}}{4G_N}\frac{1}{z_H^{d-1}}$
 \cite{Chakrabortty:2011sp}.
From (\ref{a}),the entanglement velocity can be read of  as,
\begin{equation}
 v_E= (z_h^{d-1}/z_c^{d-1})\sqrt{-h(z_c)}
 .
\end{equation}

Note that the divergence of $\alpha$ at $z_c$ is rooted in  the divergence of $\mathcal{\eta}_{\text{III}}$ integration. By expanding   $\mathcal{\eta}_{\text{III}}$ about $z=z_0$ up to the first order we get,
\begin{equation}
    \mathcal{\eta}_{\text{III}}
=
    \frac{4\pi}{\beta}
    \int_{z_H}^{z_0}
    {\frac{dz}{h(z)}}
    \left[
    1-{\left(-\frac{z_0^{2-2d}}{h(z_0)}\biggl[h(z)z^{2-2d}\biggr]_{z=z_0}'(z-z_0)
    \right)}^{-1/2}\right].
\end{equation}
 $\mathcal{\eta}_{\text{III}}$ diverges if the coefficient of $(z-z_0)$ is zero i.e, $\frac{d}{dz}\left(h(z)z^{2-2d}\right)|_{z_0}=0$ at $z_0=z_c$, that allows us to find the value of $z_c$.

We observe that the entanglement velocity shows similar qualitative behaviour as the butterfly velocity calculated in section \ref{sec:Shockandvb}, and the value of $v_E$ is always lower than the value of  $v_B$ corresponding to a fixed value of $b$. This behaviour $v_B\geq v_E$ for holographic systems was also observed in \cite{Mezei:2016zxg, Mezei:2016wfz}. Similar to the case of $v_B$, the entanglement velocity corresponding to non-zero $b$ saturates at $\sqrt{d/2(d-1)}$. Our findings agree with a variety of recent investigations, including the holographic computations, showing the universal linear growth of entanglement in quantum many body system \cite{ Liu:2013qca}.

\section{Summary and discussion}
\label{sec:summary}

In this work, we holographically study the characteristics of correlation and its degradation in the presence of chaos for a TFD state deformed by the uniform distribution of heavy and static quarks. 
The purpose of this study is twofold. Firstly, we estimate the variation of HTMI as a holographic measure of the correlation between two entangled boundary theories forming TFD state with respect to the quark backreaction. Secondly, we analyze how this correlation gets disrupted by the late time growth of an early perturbation  introduced in one of the field theories. The late time exponential behaviour of the early perturbation can also be treated as a diagnosis of chaos in the boundary theory. Using the shockwave analysis, we holographically characterises the chaos in terms of butterfly velocity and study its dependence on the backreaction parameter $b$. Moreover we compute the entanglement velocity to measure the late time behaviour of the rate of change of the correlation with respect the boundary time.


In our analysis of HTMI we find that the correlation in the boundary theory always increases with the width $l$ of the entangling region of the individual boundary. We also observe that such correlation becomes zero if the width $l$ becomes less than or equal to the critical width $l_c$. From our previous work \cite{Chakrabortty:2020ptb}, we know that the EE in each of the boundary regions increases with $b$, whereas in the present work we notice that the effect of $b$ disrupts the two sided EE. As a result we find that the TMI increases with quark deformation.

To diagnose the chaos holographically, we compute the holographic butterfly velocity and find that it gets reduced by a correction that depends on the backreaction parameter $b$. We also show how $v_B$ varies with temperature ($T$) for a particular value of $b$.
For $b=0$, $v_B$ remains unchanged with respect $T$ at a fixed value $\sqrt{d/2(d-1)}$. This result is in agreement with the butterfly velocity for AdS black hole.
For finite $b$, $v_B$ increase as $T$ also increases. Moreover, we note from figure \ref{Ivslwithoutshock} (right) that $v_B$ approaches to zero when $T \rightarrow 0$, irrespective of the choices of $b$. At high temperature, $v_B$ saturates into $v_B|_{b=0} = \sqrt{d/2(d-1)}$ for any non-zero $b$. For intermediate temperature we observe that rate of increases in $v_B$ with respect to $T$ slows down with higher values of $b$. We also investigate the HTMI in the shock wave geometry, and it is shown that as $\alpha$ increases for a particular $b$, the HTMI decreases and eventually vanishes at some value of $\alpha$. We note that the larger the value of the parameter $b$ the smaller is the value of $\alpha$ at which HTMI vanishes. We also observe that the entanglement velocity $v_E$ satisfies the constraint $v_E\leq v_B$ for any value of $b$.

We summarize the lessons we learn from our holographic analysis. 
 Firstly, the TMI becomes stronger as the value of the quark density $b$ becomes higher. Since the increase of deformation parameter $b$ enhances the number of fundamental quarks, enhancement in TMI with $b$ parameter is expected. Secondly, as $b$ increases, the exponential growth of the correlator takes a longer time to scramble the available information, thus it allows a lower value of butterfly velocity. Lastly, we expect that the  butterfly velocity  shows universal
behavior at high temperature. Moreover, TMI exhibits at both low and high temperature limits. Such universality of TMI is reported in \cite{Morrison:2012iz}. At very low temperature butterfly velocity becomes insensitive to the strength of back reaction.

 The effect of various forms of deformation and backreaction on the holographic quantum information theory and holographic quantum chaos have been previously studied in \cite{Jensen:2013lxa, Rodgers:2018mvq, Hung:2011ta, Kontoudi:2013rla, Carmi:2017ezk, Fonda:2015nma}.  Our model of backreaction is qualitatively very similar to the strongly coupled plasma consists of gluon and heavy quarks. Since, a class of strongly coupled system with well-defined holographic dual show various universal behaviors, we expect our study would be potentially useful to have a qualitative understanding of information theory and chaos in the plasma.
 
 It will be very tempting to achieve a quantitative formulation of quark backreaction in the boundary theory. Such formulation would be very insightful and leads to a better understanding of correlations in the boundary theory. We leave this analysis for the future work.

\section*{Acknowledgments}
\noindent
We thank Rudranil Basu and Rajesh Gupta for useful discussions and comments.
SC is partially supported by the ISIRD grant
9-252/2016/IITRPR/708. HH is fully supported by the ISIRD grant 9-289/2017/IITRPR/704 during a major part of this project. SP is supported by a Senior Research Fellowship from the Ministry of Human Resource and Development, Government of India. KS is supported by the institute post doctoral fellowship at IIT Bhubaneswar.



\newpage

\appendix
\renewcommand{\thesection}{\Alph{section}}
\renewcommand{\thesubsection}{\Alph{section}.\arabic{subsection}}
\renewcommand{\theequation}{\Alph{section}.\arabic{equation}}

\end{document}